\documentclass[12pt]{article}
\usepackage{amsmath,graphicx,color}
\usepackage{setspace}
\usepackage{dcolumn}
\usepackage{bm}

\begin{document}

\title{Multiply subtractive generalized Kramers-Kronig relations: application on third harmonic generation susceptibility on polysilane }
\author{\textbf{Valerio Lucarini} \\ lucarini@mit.edu \\ \\ \textbf{Jarkko J. Saarinen} and \textbf{Kai-Erik
Peiponen} \\ Department of Physics, University of
Joensuu,\\ P. O. Box 111, FIN-80101 Joensuu, Finland}
\maketitle

\date{\today}

\begin{abstract}
We present multiply subtractive Kramers-Kronig (MSKK) relations
for the moments of arbitrary order harmonic generation
susceptibility. Using experimental data on third-harmonic wave
from polysilane, we show that singly subtractive Kramers-Kronig (SSKK)
relations provide better accuracy of data inversion than the
conventional Kramers-Kronig (K-K) relations. The fundamental reason is that SSKK and MSKK relations have
strictly faster asymptotic decreasing integrands than the
conventional K-K relations. Therefore SSKK and MSKK relations can
provide a reliable optical data inversion procedure based on the use of
measured data only.

\end{abstract}

\maketitle

\section{\label{sec:level1}Introduction}
In linear optical spectroscopy Kramers-Kronig (K-K) dispersion
relations \cite{landau84} have widely been exploited in data
inversion both in absorption and reflection spectroscopy
\cite{peiponen99}. The characteristic integral structure of K-K
relations require the knowledge of the spectrum at a semi-infinite
angular frequency range. Unfortunately, in practical spectroscopy
only finite spectral range can be measured. Moreover, technical
difficulties in gathering information especially in nonlinear
optical spectroscopy on a sufficiently wide spectral range make
hard even the application of the K-K relation. The extrapolations
in K-K analysis, such as the estimation of the data beyond the
measured spectral range can be a serious source of errors
\cite{aspnes85,peiponen91}. Recently, King \cite{king02} presented
an efficient numerical approach to the evaluation of the K-K
relations. Nevertheless, the problem of data fitting is always
present in regions outside the measured range.

In the context of linear optical spectroscopy singly
\cite{ahrenkiel71} (SSKK) and multiply \cite{palmer98} (MSKK)
subtractive K-K relations have been utilized in order to avoid the
limitations due to finite-range data. Only recently they have been
proposed in the context of nonlinear optics and especially for
harmonic generation susceptibility \cite{lucarini03}. The idea
behind the subtractive K-K is that the inversion of the real
(imaginary) part of $n$th order harmonic generation nonlinear
susceptibility can be greatly improved if we have one or more
anchor points, i.e. a single or multiple measurement of the
imaginary (real) part for a set of wavelengths. Recently, we
presented generalized K-K relations and showed their validity on
polymer \cite{lucarini03b}. In this paper we present for the first
time the theory for multiply subtractive generalized K-K for the
moments of $n$th order harmonic generation susceptibility. In
addition, we show using data of polysilane that the success of
data inversion of the moments relations of third-order harmonic
generation susceptibility is better already in the case of SSKK
than for conventional K-K. The results of the latter were
presented in Ref. \cite{lucarini03b}. This SSKK is important in
the sense that usually it is rather difficult to obtain reliable
information on nonlinear susceptibility at multiple anchor points.

\section{Multiply subtractive Kramers-Kronig relations for the moments of the harmonic generation susceptibility}\label{s2}
We first present the theory related to moments of $n$th order
harmonic generation susceptibility. Then we deal with a function
$\omega^{2\alpha}\chi^{(n)}(n\omega;\omega,\cdots,\omega)$, where
$\omega$ is the angular frequency, $0\leq\alpha\leq n$, and
$\chi^{(n)}(n\omega)$ is the tensor element of the nonlinear
susceptibility. From now on we make use of the abbreviation
$\chi^{(n)}(n\omega;\omega,\cdots,\omega)=\chi^{(n)}(n\omega)$.
Next we generalize the result of Palmer $et$ $al$. \cite{palmer98}
and Lucarini $et$ $al$. \cite{lucarini03} to hold also for the
moments of the harmonic generation susceptibility. With the aid of
mathematical induction (for a rigorous proof see Appendix A in
Ref. \cite{palmer98}) we can derive multiply subtractive K-K
relations for the real and imaginary parts of as follows:
\begin{equation}\label{eq1}
\begin{split}
\omega^{2\alpha}&{\rm{Re}}\{\chi^{(n)}(n\omega)\}=\left[\frac{(\omega^2-\omega_2^2)(\omega^2-\omega_3^2)\cdots(\omega^2-\omega_Q^2)}{(\omega_1^2-\omega_2^2)(\omega_1^2-\omega_3^2)\cdots(\omega_1^2-\omega_Q^2)}\right]\omega_1^{2\alpha}{\rm{Re}}\{\chi^{(n)}(n\omega_1)\}+\cdots
\\&+\left[\frac{(\omega^2-\omega_1^2)\cdots(\omega'^2-\omega_{j-1}^2)(\omega'^2-\omega_{j+1}^2)\cdots(\omega'^2-\omega_Q^2)}{(\omega_j^2-\omega_1^2)\cdots(\omega_j^2-\omega_{j-1}^2)(\omega_j^2-\omega_{j+1}^2)\cdots(\omega_j^2-\omega_Q^2)}\right]\omega_j^{2\alpha_j}{\rm{Re}}\{\chi^{(n)}(n\omega_j)\}+\cdots
\\&+\left[\frac{(\omega^2-\omega_1^2)(\omega^2-\omega_2^2)\cdots(\omega^2-\omega_{Q-1}^2)}{(\omega_Q^2-\omega_1^2)(\omega_Q^2-\omega_2^2)\cdots(\omega_Q^2-\omega_{Q-1}^2)}\right]\omega_Q^{2\alpha}{\rm{Re}}\{\chi^{(n)}(n\omega_Q)\}
\\&+\frac{2}{\pi}\left[(\omega^2-\omega_1^2)(\omega^2-\omega_2^2)\cdots(\omega^2-\omega_Q^2)\right]{\rm{P}}\int_0^{\infty}\frac{\omega'^{2\alpha+1}{\rm{Im}}\{\chi^{(n)}(n\omega')\}}{(\omega'^2-\omega^2)\cdots(\omega'^2-\omega_Q^2)}\,{\rm{d}}\omega',
\end{split}
\end{equation}
\begin{equation}\label{eq2}
\begin{split}
\omega^{2\alpha-1}&{\rm{Im}}\{\chi^{(n)}(n\omega)\}=\left[\frac{(\omega^2-\omega_2^2)(\omega^2-\omega_3^2)\cdots(\omega^2-\omega_Q^2)}{(\omega_1^2-\omega_2^2)(\omega_1^2-\omega_3^2)\cdots(\omega_1-\omega_Q^2)}\right]\omega_1^{2\alpha-1}{\rm{Im}}\{\chi^{(n)}(n\omega_1)\}+\cdots
\\&+\left[\frac{(\omega^2-\omega_1^2)\cdots(\omega^2-\omega_{j-1}^2)(\omega^2-\omega_{j+1}^2)\cdots(\omega^2-\omega_Q^2)}{(\omega_j^2-\omega_1^2)\cdots(\omega_j^2-\omega_{j-1}^2)(\omega_j^2-\omega_{j+1}^2)\cdots(\omega_j-\omega_Q^2)}\right]\omega_j^{2\alpha-1}{\rm{Im}}\{\chi^{(n)}(n\omega_j)\}+\cdots
\\&+\left[\frac{(\omega^2-\omega_1^2)(\omega^2-\omega_2^2)\cdots(\omega^2-\omega_{Q-1}^2)}{(\omega_Q^2-\omega_1^2)(\omega_Q^2-\omega_2^2)\cdots(\omega_Q-\omega_{Q-1}^2)}\right]\omega_Q^{2\alpha-1}{\rm{Im}}\{\chi^{(n)}(n\omega_Q)\}
\\&-\frac{2}{\pi}\left[(\omega^2-\omega_1^2)(\omega^2-\omega_2^2)\cdots(\omega^2-\omega_Q^2)\right]{\rm{P}}\int_0^{\infty}\frac{\omega'^{2\alpha}{\rm{Re}}\{\chi^{(n)}(n\omega')\}}{(\omega'^2-\omega^2)\cdots(\omega'^2-\omega_Q^2)}\,{\rm{d}}\omega',
\end{split}
\end{equation}
with $0\leq\alpha\leq n$ and P denoting the principal value
integration. Here, $\omega_j$ with $j=1,\cdots,Q$ denote the
anchor points. Note that the anchor points in Eqs. (\ref{eq1}) and
(\ref{eq2}) need not to be the same. We observe that the
integrands in MSKK relations presented in Eqs. (\ref{eq1}) \&
(\ref{eq2}) have remarkably stronger asymptotic decrease, as a
function of angular frequency, than the conventional K-K
relations. The integrands of the conventional K-K relations given
for the $n$th order harmonic generation susceptibility converge
proportional to $\omega^{-(2n+2)}$ at high angular frequencies,
whereas the integrands in MSKK relations converge in relation to
$\omega^{-(2n+2+2Q)}$. Therefore, one can expect that the
limitations related to the presence of an experimentally
unavoidable finite angular frequency range are thus relaxed, and
the precision of the integral inversion is then enhanced.
According to Palmer $et$ $al$. \cite{palmer98} the anchor points
should be chosen inside the measured spectral range. Furthermore,
their study showed that accurate data inversion is possible when
the anchor points are chosen near to the zeros of the $Q$th order
Chebyshev polynomial of the first kind. In linear optical
spectroscopy it is usually rather easy to get information of the
optical constants such as real refractive index and absorption
coefficient of the medium at various anchor points. However, in
the field of nonlinear optics it may be problematic to obtain the
real and imaginary parts of the nonlinear susceptibility at
various anchor points. Therefore, in the present study we wish to
emphasize that the information even at a single anchor point
reduces the errors caused by finite data range by utilization of
SSKK. Now the choice of the location of the anchor point is not so
critical. Nevertheless, the Chebyshev zeros accumulate at the ends
of the spectral range. Therefore, we have chosen the anchor point
near one end of the data range. From Eqs. (\ref{eq1}) \&
(\ref{eq2}) we obtain for the moments of the third-order harmonic
generation susceptibility the following SSKK relations
\begin{equation}\label{eq3}
\omega^{2\alpha}{\rm{Re}}\{\chi^{(3)}(3\omega)\}=\omega_1^{2\alpha}{\rm{Re}}\{\chi^{(3)}(3\omega_{1})\}
+\frac{2(\omega^{2}-\omega_{1}^{2})}{\pi}{\rm{P}}\int_{0}^{\infty}\frac{\omega'^{2\alpha+1}{\rm{Im}}\{\chi^{(3)}(3\omega')\}}{(\omega'^{2}-\omega^{2})(\omega'^{2}-\omega_{1}^{2})}\,{\rm{d}}\omega',
\end{equation}
\begin{equation} \label{eq4}
\omega^{2\alpha-1}{\rm{Im}}\{\chi^{(3)}(3\omega)\}=\omega_{1}^{2\alpha-1}{\rm{Im}}\{\chi^{(3)}(3\omega_{1})\}
-\frac{2(\omega^{2}-\omega_{1}^{2})}{\pi}{\rm{P}}\int_{0}^{\infty}\frac{\omega^{2\alpha}{\rm{Re}}\{\chi^{(3)}(3\omega')\}}{(\omega'^{2}-\omega^{2})(\omega'^{2}-\omega_{1}^{2})}{\rm{d}}\omega',
\end{equation}
where $\omega_1$ is the anchor point.

\section{Application of SSKK relations to experimental data of third harmonic generation on polysilane}\label{s3}
We consider here the experimental data of Kishida $et$ $al$.
\cite{kishida93} related to third harmonic susceptibility of
polysilane at energy range 0.4--2.5 eV. In Figs. 1a and 2a are
shown the inverted data obtained using Eqs. (\ref{eq3}) \&
(\ref{eq4}). In the case of Fig. 1a the anchor point was selected
at the lower boundary of the data range. The data shown in Fig. 1a
for the real part of the nonlinear susceptibility has an almost
perfect match with the measured spectrum for $0\leq\alpha\leq2$,
while for $\alpha=3$ the finite range causes disagreement between
the measured and retrieved data at low energy range of the
spectrum. Nevertheless, the spectrum is well retrieved at high
energy range. In the case $\alpha=4$ the theory does not predict
convergence, but we have still reasonable agreement between
experimental and inverted curve at the high energy range. If we
compare the curves of Fig. 1a with those of Fig. 1b obtained using
the conventional K-K relations we observe that in Fig. 1b the
curve already for $\alpha=2$ is dramatically departing from the
true one at the low energy range. Thus, the SSKK relations provide
a better means than the conventional K-K relations to invert the
moments data.

In the case of the imaginary part of the nonlinear susceptibility
the anchor point was selected at a location of the low energy side
of the main spectral feature, just out of the resonance. Also in
the case of the imaginary part we observe that the SSKK procedure
provides more precise data inversion than the conventional K-K
relation (for comparison see Fig. 2b). In Fig. 2a the agreement
between measured and retrieved data is excellent for
$0\leq\alpha\leq2$ while for $\alpha=3$ we have a good agreement
except for photon energy $\leq0.7$ eV. In the theoretically
non-converging $\alpha=4$ case, we somewhat surprisingly still
have a good performance of the SSKK procedure at high energy
range. Heuristically we can interpret this result as a manifest of
the faster asymptotic decrease of the integrands realized when
SSKK relations are utilized instead of K-K relations.

\section{Conclusions}\label{s4}
We have given, as far as we know, for the first time multiply
subtractive Kramers-Kronig relations for the moments of the
arbitrary order harmonic generation susceptibility. According to
the calculations, which are based on experimental data of
polysilane, we have observed how an independent measurement of the
unknown part of the complex third-order (arbitrary-order) harmonic
generation susceptibility for a given frequency relaxes the
limitations imposed by the finiteness of the measured spectral
data. The fundamental reason is that SSKK and MSKK relations have
strictly faster asymptotic decreasing integrands than the
conventional K-K relations. Thus SSKK and MSKK relations can
provide a reliable data inversion procedure based on the use of
measured data only. That is to say no extrapolations of the data
are needed beyond the measured range. We have demonstrated that
even SSKK relations yield more precise data inversion, using a
single anchor point, than conventional K-K relations. Naturally it
is possible to exploit MSKK procedure if higher precision is
required. However, the measurement of data at multiple anchor
points may be experimentally tedious.

\section*{Acknowledgments}
The authors wish to thank Dr. Hideo Kishida and Prof. Takao Koda
for providing the measured data on polysilane, and Prof. Franco
Bassani for having suggested us to analyze this problem. Authors
J.J.S. and K.-E.P wish to express their gratitude to the Academy
of Finland for financial support. One of the authors (JJS) wishes
to thank the Nokia foundation for a grant.

\newpage 
\bibliographystyle{joeunsrt2}
\bibliography{subtractive}

\newpage

\noindent {\large{\bf{Figure captions}}}

\vspace{1cm}

\noindent {\bf{Figure 1:}} Efficacy of SSKK relation (a) versus
the conventional K-K relation (b) in retrieving
${\rm{Re}}\{\chi^{(3)}(3\omega;\omega,\omega,\omega)\}$ on
polysilane.

\vspace{1cm}

\noindent {\bf{Figure 2:}} Efficacy of SSKK relation (a) versus
the conventional K-K relation (b) in retrieving
${\rm{Im}}\{\chi^{(3)}(3\omega;\omega,\omega,\omega)\}$ on
polysilane.

\newpage
\begin{figure}
\includegraphics[scale=.85]{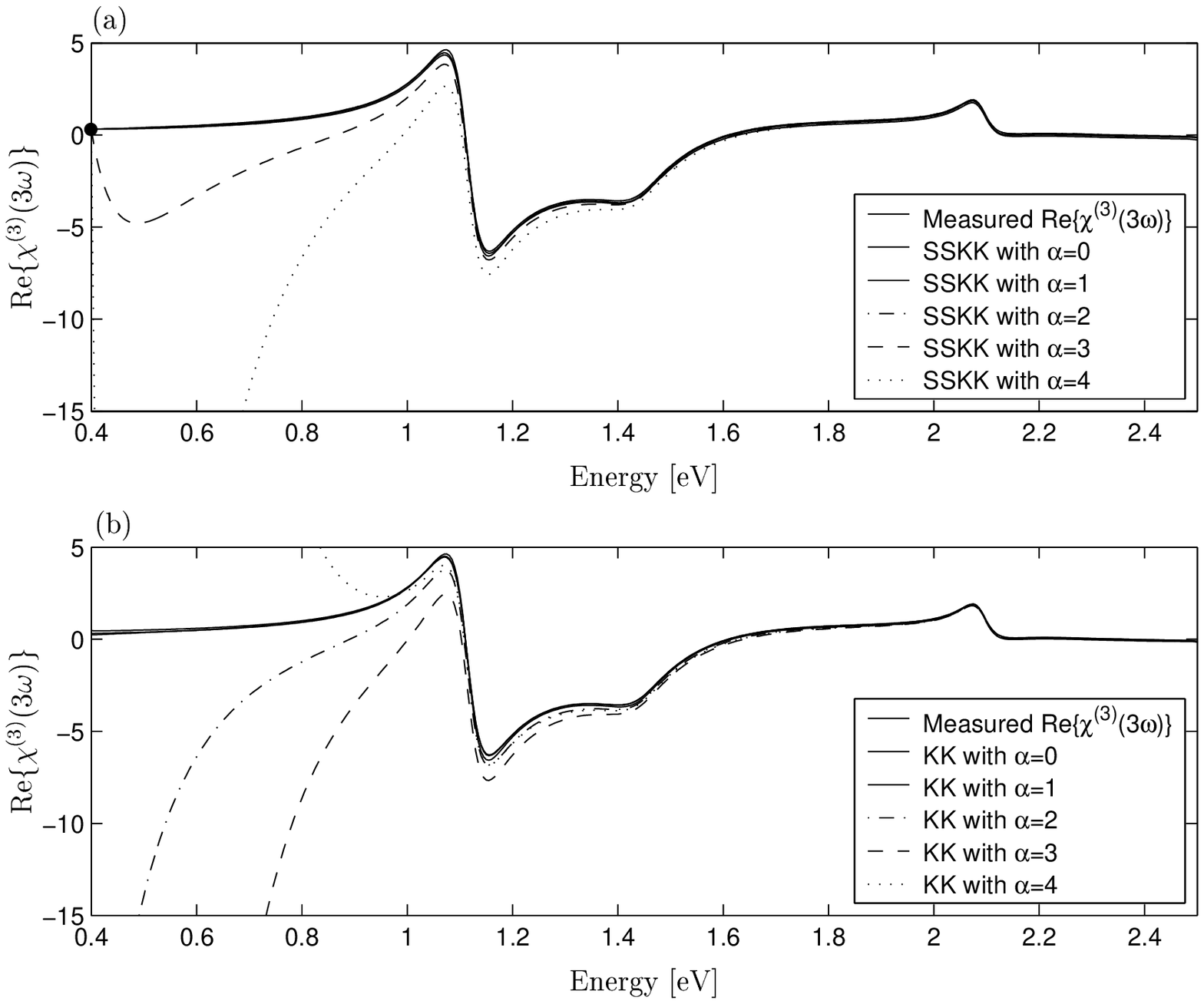}
\end{figure}
\begin{center}
{\bf{Figure 1:}} Lucarini, Saarinen, and Peiponen
\end{center}

\clearpage
\newpage
\begin{figure}
\includegraphics[scale=.85]{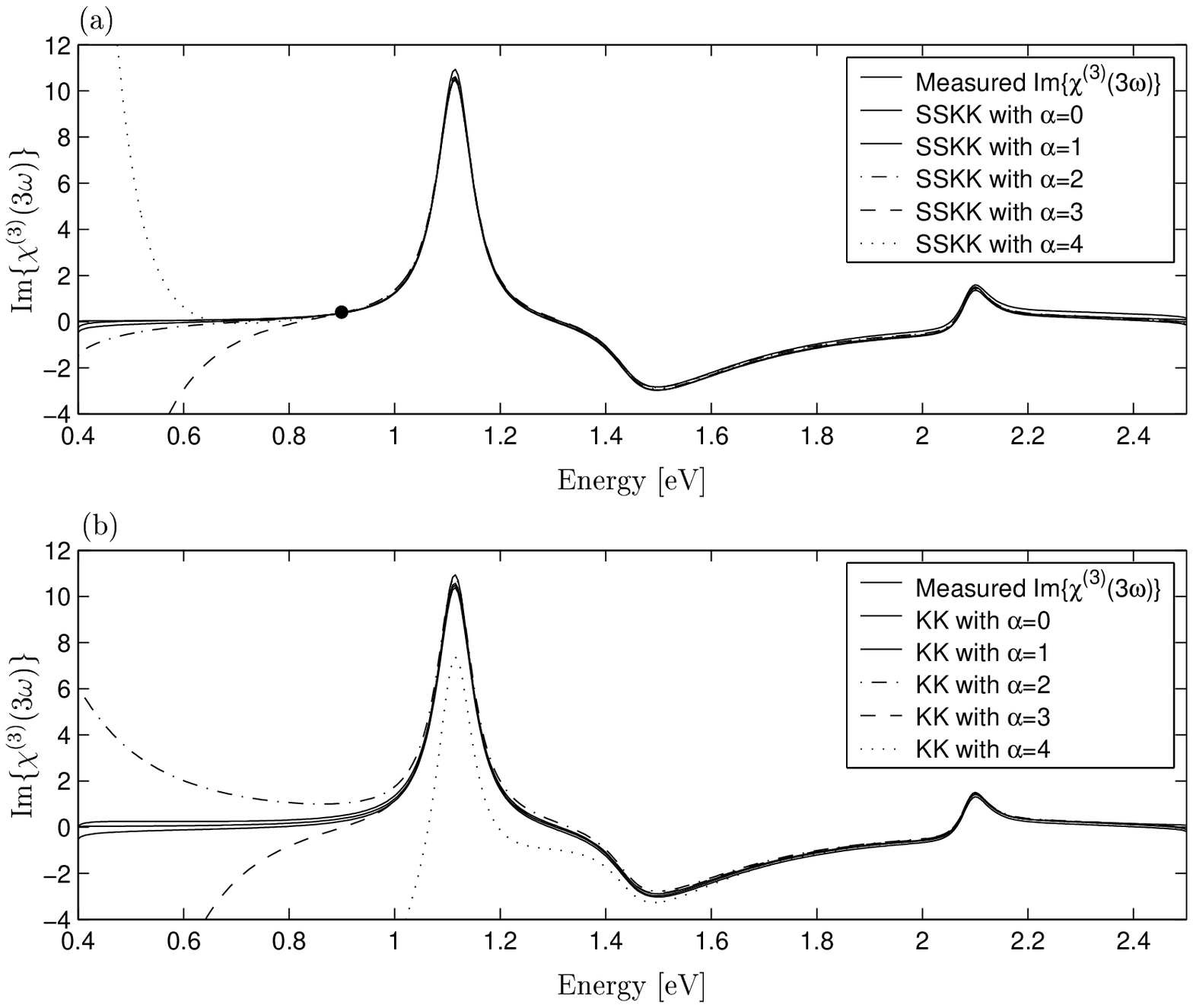}
\end{figure}
\begin{center}
{\bf{Figure 2:}} Lucarini, Saarinen, and Peiponen
\end{center}

\end{document}